\documentclass[12pt,preprint]{aastex}
\newcommand{\emp}{\rm}

\newcommand{\be}{\begin{equation}}
\newcommand{\ee}{\end{equation}}
\newcommand{\bea}{\begin{eqnarray}}
\newcommand{\eea}{\end{eqnarray}}

\newcommand{\da}{\frac{\delta}{\alpha}}
\newcommand{\di}{\partial_}
\newcommand{\ad}{\frac{\alpha}{\delta}}

\shorttitle{Halo Power-laws}
\shortauthors{Henriksen}

\begin{document}

\title{Power-laws and Non-Power-laws in Dark Matter Halos}

\author{R. N. Henriksen}
\affil{Department of Physics, Engineering Physics and Astronomy,  Queen's University, Kingston, Ontario,Canada, K7L 3N6}
\email{henriksn@astro.queensu.ca}

\begin{abstract}
Simulated dark matter profiles are often modelled as a `NFW' density profile rather than a single power law. Recently, attention has turned to the rather rigorous power-law behaviour exhibited by the `pseudo phase-space density' of the dark matter halo, which is defined dimensionally in terms of the local density and velocity dispersion of the dark matter particles. The non-power-law behaviour of the density profile is generally taken to exclude simple scale-free, in-fall models; however the power-law behaviour of the `pseudo-density' is a counter indication. We argue in this paper that both behaviours may be at least qualitatively understood in terms of a dynamically evolving self-similarity, rather than the form for self-similar infall that is fixed by cosmological initial conditions. The evolution is likely due to collective relaxation such as that provided by the radial-orbit instability on large scales.  
We deduce, from a distribution function given by first order coarse-graining, both the NFW-type density profile and the power-law pseudo-density profile.
The results are not greatly sensitive to variation about $3$ in the power of the velocity dispersion used in the definition of the phase space pseudo-density. We suggest that the power $2$ may create the more physical quantity, whose deviations from a power-law are a diagnostic of incomplete relaxation.    

\end{abstract}



\section{INTRODUCTION}

Frequently the self-similar in-fall model of structure formation is associated solely with the spherically-symmetric, power-law, purely radial, dynamics that was conclusively defined in the seminal papers by \citet{fg84} and \citet{bert85}. In such a restricted formulation, despite the non-linear exactness of the results, this model is not considered to have much application to the hierarchical-merging theory of dark matter halo formation. Moreover the model is generally thought to fail to account for the universality of form found in the simulated dark matter halos, since it predicts instead a memory of local cosmological conditions in the ultimate density power-law. 

However this memory is not in fact very far from universality. If the density of the final equilibrium halo, $\rho$, is written as a power-law in the radius $r$ in terms of the `self-similar index' (see additional discussion below) $\ad$ as $ \rho\propto r^{-2\ad}$, then the index is in fact only weakly dependent on `reasonable' cosmological conditions \citep{fg84, hw99} so that there is effective universality.
  
 This is reflected in the relation for the cosmologically fixed index \citep{fg84, hw99} $\ad=3\epsilon/(2(\epsilon+1))$, where $-\epsilon$ is the effective power of the initial cosmological peturbation and a `reasonable' value is close to $2$.  Our quantity $\epsilon$ is defined in terms of the density profile while that of \citet{fg84} is given in terms of the mass profile. This means that $\epsilon =3\epsilon_{fg}$, where $\epsilon_{fg}$ is the parameter used in \citet{fg84}. Moreover all strictly radial-orbit simulations show that the index $\ad$ tends universally to $1$ if it is initially smaller than $1$. This means that the halo density profile can never be flatter than $r^{-2}$, and it is always close to this value for $\epsilon\approx 2$. 

As an example of how the self-similar index may be coupled to standard cosmological initial conditions, we refer for example to \citet{peacock99} (p494) where the linear density profile around a maximum in the density field with power spectrum $P(k)\propto k^n$ gives $2\ad=3(3+n)/(4+n)$ or $\epsilon =3+n$. This assumes that the self-similarity is already established in this early collapse phase. The estimate works well on the galactic scale where $n\approx -1$, so that $\epsilon \approx 2$ as does $2\ad$. In \citet{hw99} the authors suggest instead $\epsilon =(3+n)/2$, which is the expected linear profile around an $n\sigma$ peak rather than about a strict maximum \citep{peebles93} (p546). This gives better results for cluster scale halos where $n\approx 1$ and $\epsilon$ is again nearly $2$. It seems that primordial clusters need not dominate their surroundings in the same way as do primordial galaxies. 

 The strict power-law behaviour is not expected to be maintained at the edge of the virialized system \citep{hw99, hld02,henrik04}. This can be due either to mass exhaustion or to tidal truncation \citep{henrik04, henrik06}. For example in a mass exhaustion situation where $\epsilon\rightarrow \infty$, the $\ad\rightarrow 3/2$ and the density profile power goes to $-3$ \citep{hw99}. The tidally truncated edge goes as $r^{-4}$ \citep{henrik04}.

It remains evident despite the preceeding caveats that the cosmologically determined self-similarity class can not be extended  all the way to the central regions of the halo , since the density there has been established as flattened relative to $-2\ad$ \citep{nfw} ( hereafter NFW), \citep{taylor01,moore98}. It seems to be the strictly radial nature of the orginal self-similar infall models that prevents them from experiencing certain kinds of relaxation \citep{mwh06,hjs,barnes05} and hence from evolving a central curved density profile.

 It is useful to recall in what follows that the self-similarity index or `class' is defined more generally in \citet{CH91} and \citet{henrik97} (see also \citet{hw95} for the steady case). In these papers it is shown to give the ratio of the length to time dimensions  (that is $L^\delta/T^\alpha$) of whatever quantity is currently controlling the self-similarity. The mass dimension is no longer independent because of the need to  maintain $G$ constant during the scaling\citep{hw95}. This formulation has the advantage of allowing the self-similar index, and thus the controlling quantity, to be a parameter in the relevant equations; which parameter  may change dynamically.

 In  \citet{hld02} and \citet{henrik04}  an asymmetric version of the self-similar infall model is treated. In general one can treat anisotropic self-similarity in both velocity and real space. Rather than solve the resulting equations by simulation, a series expansion is used in terms of the reciprocal resolution in phase space. The smallness of this parameter (essentially $1/\alpha$ with $\ad$ fixed: see details below) is related to the degree of coarse graining both in phase-space and in time. As $\alpha\rightarrow \infty$ the coarse-graining is maximal and the solution is steady. These papers have shown that it is  the maximally  coarse-grained, steady, limit of self-similar infall wherein the density behaves in simple power-law fashion. Higher order descriptions of the dynamics reveal a transitory central `flattening' ( always relative to the pure power-law, purely radial profile above) that is presumably due to collective relaxation. These terms are not always active however, being notably ineffective for the purely radial infall solution. The ultimate central density profile must arise from the dynamical evolution of the self-similar index $\ad$ (since the steady state density power-law is just the negative of twice this value, just as for radial orbits discussed above) , which phenomenon we refer to as a `running' index  in this paper.

One might protest that the running index is really a way of saying that the self-similar symmetry is lost `en route' to the relaxed state of the central regions.
However simulations show an indication of an underlying self-similarity in other ways. One indication is the proportional growth of the virial radius and the NFW scale radius (i.e. constant NFW `concentration') \citep{mwh06}). Another indication is the power-law behaviour of the dimensional or pseudo phase-space density (the real density being the distribution function or DF) \citep{taylor01,dehnen05,barnes05, barnes06,austin05}. If then the scaling symmetry is broken during the dynamical relaxation of the central halo, it should only be weakly broken in the sense that we can expect a solution in the form of a perturbed self-similarity. 
 If the evolution is slow relative to a particle crossing time, then at each epoch the system may be regarded as self-similar with the current similarity index. The running index thus corresponds to a kind of `adiabatic' self-similarity.It is precisely in this form that we seek a solution for the DF to first order.  

So holding the similarity index strictly constant, as is done in the radial self-similar infall models, does not  allow for effective internal relaxation. There are formal higher order terms available for the radial self-similar infall \citep{hld02} which in principle describe the approach to a relaxed state, but  they do not arise in the absence of a suitable mechanism. This has been explicitly confirmed in the simulations by \citet{mwh06}. Analytically, as is emphasized in section 4.1 of this paper, a radially biased DF (correct to first order) does not allow a curved density profile to coexist with a power-law pseudo-density. Such coexistence  is however characteristic of the different self-similarity in the halo centre. The purely radial-shell instability detected in \citet{hw97} for the original models is thus not the effective mechanism relaxing the halo, which is consistent with the moderate redistribution of energy found in these cases \citep{hw99}.  

In \citet{hw95} a class of spherically symmetric, stationary, isotropic power-law solutions permitting flatter profiles than the cosmologically fixed radial infall models was found. These solutions include the \citet{EC97} profile and the NFW profile \citep{henrik06}, which the cosmologically fixed radial self-similarity can not do. Simulations show \citep{hjs,mwh06} that such solutions are closer to the state of  the halo centres, although the envelopes are closer to the cosmologically fixed radial self-similar infall \citep{hw99,lu} until edge effects are encountered. Once again a `running' self-similar index is implied, but such steady  solutions do not describe the temporal approach to equilibrium during which the relaxed state is established. This requires the higher order terms in our series expansion. 
 
As a result of these preliminary considerations, two questions are to be addressed in this paper:
 
(i) Does the `similarity class' (the value of the parameter $\ad$ ) evolve dynamically during the halo growth due to collective relaxation in anisotropic systems, and 

(ii) can this evolution be such as to destroy the power-law density profile while maintaining a power-law pseudo-density and constant concentration? 

The answer to the first of these questions was implicitly assumed to be yes in the discussion of \citet{henrik06}, since there it was argued that the central similarity class of an isolated core could be set by conserving particles and phase-space volume rather than by the cosmological outer conditions.  A `running' self-similar index is in fact compatible with the formulation of self-similarity pioneered in \citet{CH91} (see also \citet{henrik97} and comment below).

In this paper therefore we explore the predictions of the first order term in the coarse-graining series of the self-similar system \citep{hld02}. This term is time dependent and includes in a transitory fashion (until it becomes too large) the approach to the equilibrium state due to collective relaxation.  In addition the dynamical system is regarded as being adiabatically self-similar so that the index $\ad$ is free to vary from a value slightly greater than $1$  (where it is set by reasonable scale free cosmological conditions \citet{fg84} and above) to a value approaching $0.5$ or less  as the density profile power passes through $-1$. It is important to understand however that before this ultimate steady state is reached, the first order terms can curve the density profile in such a fashion as to fit the simulations over a limited radial range. Our conclusion wll be that these same terms do not perturb the pseudo phase-space density so strongly in the same radial range.

In the next section after a brief review  of the method employed, we calculate the first order coarse-grained terms for spherically symmetric self-similarity with velocity anisotropy \citep{hw95, hld02,henrik04}. The spherical symmetry should not be thought of as an essential limitation since similar equilibrium solutions exist with general symmetry \citep{henrik04} (appendix A).  Spherical symmetry does however allow additional integrals of the motion which increase the choices available for the DF \citep{kl92,hw95,henrik04}. Since there are no explicit non-radial forces in spherical symmetry, we are forced to manually reproduce the evolution of the DF from radial orbits to isotropy. The increased choice of DF in spherical symmetry permits this. Thus we are able to choose a core DF in section (4.2) that corresponds to the spherically averaged equilibrium DF, which form persists even in general symmetry \citep{henrik04}. 

Section 3 discusses the calculation of the pseudo phase-space density (referred to frequently as `pseudo-density' for brevity)  in terms of the higher order DF, and subsequently detailed results for several DFs are presented in a section 4. Section five draws our conclusions and gives a brief speculative discussion.

\section{Multi-dimensional self-similarity}

\subsection{Summary of the Lie symmetry description of self-similarity}

In the mathematical description of self-similarity introduced in \citet{CH91} and applied to the joint Poisson-Boltzmann system in \citet{henrik97} (with spherical symmetry and anisotropic velocity space), one introduces a scaling vector of the form 
\be
{\bf b}\equiv (\alpha,\delta,\nu,\lambda,\mu).
\ee

The quantities $\alpha$, $\delta$, $\nu$, $\lambda$, $\mu$ give the logarithmic (additive) form of any scaling in the anisotropic phase space of spherically symmetric dynamics, namely, (time, distance, velocity,$ j^2$, mass). The symbol $j^2$ stands for mean square specific angular momentum. This  set is complete for dynamical purposes in the sense that the dimension of any physical quantity $Q$ may be expressed in terms of its co-vector $ {\bf d}_Q$ in this dimension space. For example the dimensions of Newton's constant $G$, namely (in obvious notation) $L^3/(MT^2)$, can be expressed as the dimensional co-vector ${\bf d}_G=(-2,3,0,0,-1)$. This is not a unique representation, but since the other dimensions can be expressed in terms of time, length and mass; it is the natural one and all other representations are equivalent. 

This description of the self-similar symmetry is that of a Lie symmetry in phase-space along the scaling direction ${\bf k}$ (see \citet{CH91} for a more precise mathematical description) so that the change in any quantity $Q$  is 
\be
{\bf k}{\cal L}Q=({\bf b}\cdot {\bf d}_Q)Q,\label{Q}
\ee
and the Lie differentiation is given in spherical symmetry by 
\be
{\bf k}{\cal L}\equiv \alpha t\partial_t+\delta r\partial_r+\nu v_r\partial_{v_r}+\lambda j^2\partial_{j^2}+\mu m\partial_m.
\ee
One proceeds to make the differentiation along ${\bf k}$ easy by defining a logarithmic time variable $T$ satisfying
\be
\partial_{x^i}T=0,~~~~~\alpha t\partial_t T=1,\label{T}
\ee
and a set of variables, $X^i$, which are invariant under the Lie scaling symmetry so that they satisfy
\be
({\bf k}\cdot\partial_{\bf x})X^{j}=0.\label{XYZ}
\ee

In these equations ${x^i}$ are the set of physical variables ($(r,v_r,j^2)$ in this example) and the set $X^i$ of the same number are found as characteristic constants of the linear partial differential equation (\ref{XYZ}). 
The $X^i$  constitute the self-similar variables and this procedure defines `multi-dimensional' self-similarity.

The important feature of the self-similar variables is that when these are used to describe the system the scaling operation ${\bf k}{\cal L}Q$ becomes simply $\partial_T$, and so equation (\ref{Q}) on integration yields the multiplicative scaling 
\be
Q=Q_o(X^j)~e^{({\bf b}\cdot {\bf d}_Q)T}.\label{multiQ}
\ee
The function $Q_o$ is independent of $T$ when the scaling symmetry holds, but if there is no such symmetry than it will be in general  also a function  of $T$. In such a case the variables $X^i$ are simply a useful set, alternate to the $x^i$.

In the present example equation (\ref{XYZ}) gives by the method of characteristics the set {$X^i$} as (we use notation consistent with previous work)
\bea
R&\equiv& r~e^{-\da(\alpha T)},\nonumber\\
Y&\equiv& v_r~e^{-(\nu/\alpha)\alpha T},\label{SSvars}\\
Z&\equiv& j^2~e^{-(\lambda/\alpha)\alpha T},\nonumber
\eea
 and the logarithmic time is 
\be
e^{\alpha T}=\alpha t.\label{expT}
\ee

There are two additional subtleties of the method. Because for a self-gravitating system $G$ is fixed under a scaling operation, we must have ${\bf b}\cdot {\bf d}_G=-2\alpha+3\delta-\mu=0$ by equation (\ref{multiQ}). Hence the mass scaling is no longer independent of the space and time scaling, becoming rather $\mu=3\delta-2\alpha$. The ${\bf b}$ vector can then be reduced by one dimension. Thus for example the mass distribution function $f(r,v_r,j^2)$ has ${\bf d}_f=(3,-6,0,0,1)$, which on eliminating $\mu$ according to the preceeding relation becomes $(1,-3,0,0)$. Then equation (\ref{multiQ}) allows us to write (the factor $\pi$ is used for convenience)
\be
\pi f(r,v_r,j^2)\equiv F(r,v_r,j^2)=P(R,Y,Z)~e^{-(3\da-1)\alpha T},\label{DF}
\ee
where $P$ would also depend on $T$ if there were no multi-dimensional self-similarity. 

 In like manner, velocity and angular momentum scalings can be written in terms of $\alpha$ and $\delta$ as $\nu=\delta-\alpha$ and $\lambda = 4\delta-2\alpha$. However since there is no invariant velocity in the problem $\nu$ remains free so that $\delta\ne \alpha$, and similarly $4\delta\ne 2\alpha$ since there is no invariant angular momentum in the problem and $\lambda$ remains free (near the edge of the system there may be tidally induced angular momentum which alters this latter remark \citep{henrik04}, as would the presence of a fixed velocity $c$ alter the conclusion for $\nu$).   

 We see therefore that the parameters $\alpha$ and $\delta$  suffice to express the dimensions of any dynamical quantity in our problem, so that from now on we drop the $\nu$ and $\lambda$ components from ${\bf b}$ in addition to the $\mu$ component as discussed above. Moreover it is only the ratio $\da$ or equivalently $\ad\equiv a$ that appears as a parameter in our transformed equations. This parameter $a$ is generally left undetermined so that either appropriate initial conditions, or such conditions as arise dynamically, may determine the self-similarity. Following \citet{CH91} $a$ is referred to as the `similarity class' since it now determines completely the self-similar symmetry.

The second subtlety is the one introduced in \citet{hld02}. The phase space volume is transformed under the transformation to self-similar variables as
\be
\Delta r~\Delta v_r~\Delta j^2 =\Delta R~\Delta Y~\Delta Z~e^{(6/a-3)\alpha T}.\label{resolution}
\ee
Thus letting $\alpha\rightarrow\infty$ gives an ever coarser-graining of the self-similar phase space ($a$ is held constant) for fixed increments in the ${R,Y,Z}$ at fixed $T$, while in addition giving the  temporal asymptotic limit by equation (\ref{expT}).  Our method of coarse-graining consists in solving the
transformed equations (expressed in terms of the self-similar variables) in an inverse series in $\alpha$. The zeroth order gives the steady, equilibrium, limit, but the higher order terms can describe the approach to this limit in finer detail. By interpreting the coarsest grained solution as the equilibrium solution, we are relying on the notion that a fully relaxed system should show the same DF at different resolutions, and that this condition is reached after an indefinitely long time.  

\subsection{Formulation in spherical symmetry with velocity anisotropy}

 In this subsection we follow the treatment of \citet{hld02} while making explicit the units of the various quantities. 
We recall in our notation that $v_r$ is the radial velocity and that the specific angular momentum is $j^2\equiv r^2 v_\perp^2$, with $v_\perp$ the transverse velocity on a sphere. We also use the logarithmic time of equation (\ref{expT}) and the method of the preceeding section.

It is important to distinguish units from dimensions in this method. We do not assume `a priori' essential constants with dimensions of velocity, length, or density, whose existence would constrain the scaling symmetry as noted in the previous section. However we do pick arbitrary units for these quantities, sometimes called fiducial values. Our various quantities are consequently dimensionless numbers with these units understood. 
 
The unit of the DF is $F_o$ while that of time, radius, velocity and density are $r_o/v_o$, $r_o$, $v_o$ and $\rho_o$ respectively. To obtain the equations in the form that we use, these units are related by 
\bea
  ~~~~~~F_o&=&\rho_o/v_o^3\nonumber\\
~~~~~~~~~~~& &~~~~~~~~~~~~~~~~~ \\
 ~~~~~~v_o^2&=&4\pi G\rho_o r_o^2.\nonumber
\eea
The potential is measured in units of $v_o^2$.

The self-similar variables found by the procedure of the previous section are (we repeat equation (\ref{SSvars}) for convenience)
\be 
R\equiv r e^{-\da(\alpha T)},~~~~~ Y\equiv v_re^{-(\da-1)\alpha T}~~~~Z\equiv j^2e^{-(4\da-2)\alpha T}.\label{variables}
\ee
 The remaining dependent variables namely density $\rho$ and potential $\Phi$ are expressed according to equation (\ref{multiQ}) using the reduced vector ${\bf b}$. These can be written as  
\be
\rho(r,t)\equiv \Theta(R) e^{-2\alpha T}~~~~~ \Phi(r,t)\equiv \Psi(R,T) e^{2(\da-1)\alpha T}.\label{rhophi}
\ee
Together with the dependent variable $P(R,Y,Z)$ of equation (\ref{DF}), $\Theta$ and $\Psi$ are the dependent variables that express multi-dimensional self-similarity. 
As remarked in the previous section  the transformation to these variables does not imply self-similarity unless we set $\partial_T=0$ in the Boltzmann equation. Otherwise they remain perfectly general, and $P$, $\Theta$ and $\Psi$ would also depend on $T$. 

The concept of `running' or adiabatic self-similarity discussed in the introduction is compatible with our formulation (that is equation (\ref{XYZ}) may still be used to find the self-similar variables) if $\delta$ in the equations is replaced by a time dependent quantity 
\be
\delta_a\equiv (1/T)\int^T~\delta(T')dT',\label{running}
\ee
while the coarse-graining parameter $\alpha$ is left unchanged. Moreover $\partial_T$ acting on $P$ must be considered small  during a dynamical time. Then one requires $\delta_a$ to nearly scale with $\alpha$ during the coarse-graining as usual to preserve approximately the similarity class $a=\alpha/\delta_a$, but it must retain sufficient variation in $T$ relative to $\alpha$ to permit $\ad$ to change by about a factor $2$ or less between the envelope and core in the application to dark matter halos. We shall not in fact need this formulation explicitly in this paper since it suffices simply to consider different $\ad$ in the core and envelope, which procedure corresponds to our similar treatment of the evolving DF by fixing the end points. The function $\delta_a(T)$ contains all of the internal relaxation physics, and hence is not easily known.

With these definitions the Poisson equation and the collisionless Boltzmann equation (CBE henceforth) become respectively
\be
\frac{1}{R^2}\di R(R^2\di R\Psi)=\Theta,
\label{Poiss}
\ee
and 
\bea\label{CBE}
\frac{1}{\alpha}\di T P&-&(3\da-1)P+(\frac{Y}{\alpha}-\da R)\di R P\nonumber\\
&-&\left((\da-1)Y+\frac{1}{\alpha}(\di R\Psi-\frac{Z}{R^3})\right)\di YP-(4\da-2)Z\di ZP=0,
\eea
where the scaled density and DF are related by
\be
\Theta=\frac{1}{R^2}\int~P~dY~dZ.
\label{density}
\ee

\subsection{Self-similar relaxation}

We proceed as usual  by writing a series for $P$ and $\Psi$ in inverse powers of $\alpha$ while holding the similarity class constant \citep{hld02,henrik04}, which we write consistently as $a$ below. We will follow the time dependence of the similarity class only by letting $a$ be a variable parameter so that $\di TP=0$ in the equations. Then one finds by retaining terms to second order in the governing equations that the solution for $P$ becomes \citep{hld02}
\be
P=P_{oo}(\zeta_1,\zeta_2^2)R^{a-3}-\frac{1}{\alpha}P_{11}(\zeta_1,\zeta_2^2)R^{-3}+\frac{1}{2\alpha^2}P_{22}(\zeta_1,\zeta_2^2)R^{-(3+a)}
\label{second}
\ee
where the variables $\zeta_1$ and $\zeta_2^2$  are defined as 

\be
\zeta_1\equiv YR^{(a-1)},~~~~~\zeta_2^2\equiv ZR^{2(a -2)}
\label{C-variables}
\ee
and are constants on the characteristics of the CBE at all orders of the expansion. The functions $P_{ii}$ are only known as solutions to $i+1$ equations when the series is terminated at the $(i+1)^{th}$ order, as we describe further below.  

The zeroth order (maximally coarse-grained and steady) density and potential become respectively (no black hole) using equation (\ref{Poiss}):
\be
\Theta_o=R^{-2a}I_{oo},~~~~I_{oo}\equiv \int~P_{oo}~d\zeta_1~d\zeta_2^2,
\label{dzero}
\ee
and (for $a\ne 1$, which case is logarithmic) 
\be
\Psi_o=-\gamma_o R^{2(1-a)}, ~~~~~\gamma_o\equiv \frac{I_{oo}}{2(a-1)(3-2a)}.\label{zpot}
\ee
Note that the quantities $I_{oo}$ and $\gamma_o$ are dimensionless numbers.

As an illustration of the physical implication of these latter relations, we note that equations (\ref{variables}),(\ref{DF}) and (\ref{rhophi}) reveal that to zeroth order $\pi f_o=P_{oo}r^{(a-3)}$, $\rho_o=I_{oo}r^{-2a}$ and $\Phi_o=-\gamma_o r^{2(1-a)}$. This constitutes a steady solution since $\zeta_1$ and $\zeta_2^2$ are also seen to be independent of time, based on the definitions in equation (\ref{variables}). 

 The function $P_{11}$ is given formally by \citet{hld02} as corrected in \citet{henrik04} as:
\be
P_{11}\equiv \left((a-1)(\zeta_1^2-2\gamma_o)+\zeta_2^2\right)\di {\zeta_1} P_{oo}+2(a-2)\zeta_1\zeta_2^2\di {\zeta_2^2}P_{oo}-\zeta_1(3-a)P_{oo},
\label{P11}
\ee
and we find  by extending the series to the second order the function $P_{22}$ to be
\be
P_{22}\equiv ((a-1)(\zeta_1^2-2\gamma_o)+\zeta_2^2)\di {\zeta_1}P_{11}+2(a-2)\zeta_1\zeta_2^2\di {\zeta_2^2}P_{11}-3\zeta_1P_{11}+(3a-2)\gamma_1\di {\zeta_1}P_{oo}.
\label{P22}
\ee
The first order potential is given from (\ref{Poiss}) as (for $a\ne2/3$, which case is logarithmic)
\be
\Psi_1=-\gamma_1R^{(2-3a)}, ~~~~~\gamma_1\equiv \frac{I_{11}}{3(1-a)(2-3a)},
\label{psi1}
\ee
where 
\be
I_{11}\equiv \int~ P_{11}~d\zeta_1~d\zeta_2^2.
\label{I11}
\ee

The nature of the series (\ref{second}) requires some comment here. It is evident that for any finite $\alpha$ the series risks to be non-convergent as $R\rightarrow 0$, depending on the nature of the functions $P_{11}$ and $P_{oo}$. Stopping the series at a chosen order by rendering all higher order terms zero as in \citep{henrik04,hld02}, will determine these functions (see discussion to follow) so as to determine an optimum transitory expression for the DF to the required order. However there will always be an inner limit in $R$ to the validity (except as $\alpha \rightarrow \infty$, which is the completely relaxed steady state) whenever the relevant $P_{ii}$ do not vanish with $R$, and this translates at any $r$ to an upper limit in $t$. In the non-spherically symmetric case, where in principle all of the relaxation physics is included, it may be worth performing a renormalization group improvement to the correction term as outlined in 
\citet{McD06}.

The zeroth order solution is determined in velocity space by requiring $P_{11}=0$ (\citet{hld02,henrik04}), and this yields directly by the method of characteristics on what is now equation (\ref{P11})
\be
P_{oo}=K(\kappa){\cal E}^{\frac{3-a}{2(a-1)}},
\label{zeroDF}
\ee
where 
\be
\kappa\equiv {\cal E}(\zeta_2^2)^{\frac{a-1}{2-a}},
\label{kappa}
\ee
where 
\be
{\cal E}\equiv |\frac{\zeta_1^2+\zeta_2^2}{2}-\gamma_o|.
\label{epsilon}
\ee
Here $K$ is an arbitrary function of $\kappa$.

The special case $a=1$ is exponential (\citet{henrik04}) and formally describes the singular isothermal sphere, but we do not pursue it here since we believe it to be unstable to the development of a core.

Once again we note that we can pass to physical variables if we wish, by using the various definitions and transformations above. These show that 
\be
\pi f_o=K(\kappa)|E|^{(\frac{3-a}{2(a-1)})},~~~~~ {\cal E}=|E|r^{2(a-1)},\label{zeropDF}
\ee
and 
\be
|E|=|\frac{v_r^2}{2}+\frac{j^2}{2r^2}+\Phi|,~~~~~ \kappa=|E|(j^2)^{(\frac{a-1}{2-a})}.
\ee 
This order is the steady, equilibrium, limit as $\alpha\rightarrow\infty$ and was originally derived in \citet{hw95} although it had been already postulated in a specific form by \citet{kl92}. It is interesting that a proper choice of similarity class $a$ allows one to include the DF's associated with most commonly used density profiles in their power-law limits. Thus referring to the work of \citet{Tretal} in the high energy, small $r$, limit (that is for $a>1$: it is the low energy, small $r$ limit for $a<1$) and identifying their parameter $\eta$ with our $3-2a$, one infers from their equation (19) the same DF as in equation (\ref{zeropDF}) for $3/2>a>0$. The particular case $a=1/2$ yields the central NFW power-law density profile and approximate DF. \citet{henrik06} has suggested how this might arise dynamically  due to conserved phase-space volume in an isolated core.

In the present work we are concerned with going beyond this limit to the first order time-dependent corrections in the DF and the density profile. We will then use the corrected DF to calculate the pseudo-density profile in order to test whether this quantity is more or less sensitive to the correction. We might hope to find a flattening central density coexisting with an accurate power-law in the pseudo phase-space density.

In order to carry out this plan we need the functions $P_{oo}$ and $P_{11}$. By stopping the series at first order we require $P_{22}=0$.  Equation (\ref{P11}) and $P_{22}=0$  are then the expected two coupled partial differential equations for the functions $P_{oo}(\zeta_1, \zeta_2^2)$ and $P_{11}(\zeta_1,\zeta_2^2)$.

These two equations may be regarded as separate quasi-linear equations with a source function that depends on the other unknown function. They have the same characteristics in the zeta space namely ($\ell$ is measured along a characteristic);
\bea
\frac{d\zeta_1}{d\ell}&=&(a-1)(\zeta_1^2-2\gamma_o)+\zeta_2^2,\\
\frac{d\zeta_2^2}{d\ell}&=&2(a-2)\zeta_1\zeta_2^2,
\eea

and these yield as a first integral the same characteristic constant as that given in equation (\ref{kappa}) above. But the characteristic equations for the functions themselves become
\be     
\frac{dP_{oo}}{d\ell}=P_{11}+(3-a)\zeta_1P_{oo},
\label{ChPoo}
\ee
and 
\be
\frac{dP_{11}}{d\ell}=3\zeta_1P_{11}+\frac{I_{11}}{3(1-a)}\left(\di {\zeta_1}P_{oo}\right)_{char}.
\label{ChP11}
\ee
The difficulty presents itself in the last term of this latter equation wherein $P_{oo}$ must be known generally in order to evaluate the required derivative on the characteristic in zeta space. The only reasonable resolution is by iteration.

We proceed by substituting the zeroth order expression (\ref{zeroDF}) for $P_{oo}$ in equation (\ref{ChP11}) and solving for the next approximation to $P_{11}$ by the method of characteristics to find ($a\ne 2/3$)
\be
P_{11}=-sgn(a-1)\gamma_1K(\kappa)({\cal E})^{\left(\frac{5-3a}{2(a-1)}\right)}(\frac{3-a}{2(a-1)}+\frac{d\ln{K}}{d\ln{\kappa}})+K'(\kappa)({\cal E})^{\frac{3}{2(a-1)}}.
\label{itP11}
\ee
The function $K'$ is arbitrary as is also, we recall, $K$. One might use this last result to find the next order correction for $P_{oo}$ from equation (\ref{ChPoo}), but this will be a small correction and will be ignored in what follows. This term could now be transformed into physical variables as was done above for the zeroth order, but we shall not need this except to note that unlike the zeroth order, the first order is not time independent. By stopping at second order in this way the whole series is terminated; but for terms of the same form as the zeroth order, which may be absorbed into a renormalized zeroth order term. 

We are now able to calculate directly the density and phase-space pseudo-density behaviour to first order for various choices of the functions $K$ and $K'$, which must be adopted on the basis of physical considerations. We proceed to identify the form of these corrections in the next sectio and the nature of our choices of DF in the following section.

\section{Density and phase-space pseudo-density to first order}

The density behaviour to first order is readily calculated from equation (\ref{density}), together with the various definitions and the first order series  to be 
\be
\frac{\rho r^{2a}}{I_{oo}}=\left(1-\frac{I_{11}}{\alpha I_{oo}}R^{-a}\right ).
\label{1stden}
\ee

This expression shows that the density can `flatten' relative to its outer slope $-2a$ near the centre of the system for two reasons. In the first case \citep{hld02} the second term in brackets on the right increases as $R\rightarrow 0$, and over a limited range (see e.g. figure \ref{fig:curvefit}) the right hand side combines with the power on the left hand side of the equation to yield a  density profile close to the NFW profile. The active term on the right increases in time at fixed $r$ however and so it gives  a transitory effect at fixed $r$, since the density may not become negative. Explicitly we must have $R^a>I_{11}/(\alpha I_{oo})$, which with the definition of $R$ becomes $r^a/t>I_{11}/I_{oo}$. Hence for any fixed $a$ and $r$ there is an upper limit to $t$ for which the first order term may be used (see also figure \ref{fig:curvefit} where $t$ is fixed and the minimum $r$ is found for validity of the first order). This is true so long as $I_{11}/I_{oo}$ is finite, which is the case except for the limiting values $a=1$ and $a=2/3$. This ratio can be quite small near these values where it is essentially linear in $a$. 
 
The  steady state is found by letting $\alpha\rightarrow \infty$, which removes the flattening term on the right, and so the second or ultimate flattening must be due to the `running' nature of the self-similarity index $a$ as it decreases through unity to $0.5$ or less , as expressed by the power on the lhs. The puzzle that we address here is how can the phase-space pseudo-density avoid these deviations from a power-law?
 
\citet{dehnen05} show that the pseudo-density based on the radial velocity dispersion $\sigma$ satisfies as good or better a power-law as that based on the total velocity dispersion, so we shall use this  for simplicity. In \citet{hansen04} Hansen has studied the consequences of assuming (both strict and approximate) power-law behaviour for the density and the pseudo-density together. He has independently deduced similar results to our findings below (for example $a=1/2$ is a limiting case) using $\rho/\sigma^\epsilon$ for the pseudo-density. We define the pseudo-density similarly, but with $\epsilon=3e$ for notational reasons.

We calculate $\rho/\sigma^{3e}$ first rather formally from our various definitions and find (we expect $e$ to be of order unity)  :
\be
\frac{\rho}{\sigma^{3e}}r^{((2-3e)a+3e)}=\frac{R^{2a}\Theta}{(\overline{\zeta_1^2})^{3e/2}},
\ee
and so using explicitly our expansion to first order in (\ref{density}) and the definition of $\sigma$ one obtains
\be
\frac{\rho}{\sigma^{3e}}r^{((2-3e)a+3e)}\frac{M_{oo}^{3e/2}}{I_{oo}^{(1+3e/2)}}=\frac{\left(1-\frac{I_{11}}{\alpha I_{oo}}R^{-a}\right)^{(1+3e/2)}}{\left(1-\frac{M_{11}}{\alpha M_{oo}}R^{-a}\right)^{3e/2}}.
\label{pseudo}
\ee

In this expression the integrals $I_{oo}$ and $I_{11}$ have already been defined above in eqs. (\ref{dzero}) and (\ref{I11}), and we have introduced 
\be
M_{oo}\equiv \int~\zeta_1^2~P_{oo}~d\zeta_1~d\zeta_2^2,
\label{Moo}
\ee
and
\be
M_{11}\equiv \int~\zeta_1^2~P_{11}~d\zeta_1~d\zeta_2^2.
\label{M11}
\ee

We observe from eq. (\ref{pseudo}) that if the right-hand side is roughly constant with $R$, then the power-law predicted for the phase-space pseudo-density is $3e(a-1)-2a$, that is just greater than $-2$ for $a=1+$ and $e=1$. As also noted in \citet{hansen04}, this slope is $-2$ independent of $a$ when $e=2/3$, and it is approximately $-2$ independently of $e$ when $a\approx 1$. It is clear that the right-hand side is not strictly constant however, but the question is rather; can it be more slowly varying than the correction factor to the density on the right of eq. (\ref{1stden})? 

The answer to the previous question requires a tedious calculation of the integrals in expression (\ref{pseudo}) for reasonable choices of the DF, which will be the subject of the next section. We note here that if the first order terms in equation (\ref{pseudo}) are small (they must be less than unity), then by first order expansion of the terms in brackets, our calculations must show that 
\be
 \frac{\rho}{\sigma^{3e}}r^{((2-3e)a+3e)}\frac{M_{oo}^{3e/2}}{I_{oo}^{(1+3e/2)}}\approx 1+\frac{I_{11}R^{-a}}{\alpha I_{oo}}\left(\frac{3e}{2}\left(\frac{I_{oo}M_{11}}{I_{11}M_{oo}}-1\right)-1\right)
\label{linear}
\ee
is slowly varying compared to the density. This requires the quantity in exterior brackets on the right of this latter expression to be small, which we shall refer to as the `correction factor' (C) for brevity, so that 
\be
C\equiv \frac{3e}{2}\left(\frac{I_{oo}M_{11}}{I_{11}M_{oo}}-1\right)-1.
\label{C}
\ee 

This approximation serves for our present purposes, but there is no reason in principle why the more exact expression (\ref{pseudo}) should not be used. This is so provided that the general question is also formulated in such a way that any constants, which depend on particularities of the system such as mass, cancel out. This may be achieved by calculating the ratio of the logarithmic derivatives  of $PsD\equiv (\rho/\sigma^{3e}) r^{((2-3e)a+3e)}$ and $D\equiv \rho r^{2a}$ respectively
to obtain 
\be
rlogds\equiv d\ln{(PsD)}/d\ln{R}\times d\ln{R}/d\ln{D}=-\frac{(C+\frac{I_{11}R^{-a}}{\alpha I_{oo}}\times \frac{I_{oo}M_{11}}{I_{11}M_{oo}})}{(1-\frac{I_{11}R^{-a}}{\alpha I_{oo}}\times \frac{I_{oo}M_{11}}{I_{11}M_{oo}})},\label{PsDD}
\ee
where we may use equation (\ref{C}) to write 
\be
\frac{I_{oo}M_{11}}{I_{11}M_{oo}}\equiv 1+\frac{2}{3e}(1+C).\label{1stratio}
\ee

In this approach we would look for a small value of this ratio as a function of $a$, $e$ and $r$ as an idicator of good power law behaviour in the pseudo-density even as $D$ becomes a significant function of $r$. An advantage of this approach is that we do not have the functions of $I_{oo}$ and $M_{oo}$ on the left hand side of equation (\ref{pseudo}) to consider, for although these are formally constant in $r$ they do depend on $a$ and so implicitly on $r$ as $a$ `runs'. This is only a weak effect when the first order terms are small however and we shall find numerically below that ignoring this variation in the linear approach does not yield gross error.
It is in fact already evident from equation (\ref{PsDD}) that putting $C=0$ renders the exact ratio of the logarithmic derivatives as a small quantity of first order.

\section{Distribution Functions and Integrals}

We look at various cases of the DF in order to see when $C$  of equation (\ref{linear}) may  indeed be small.

\subsection{Envelope-Core transition region}
 
We know that the orbits in the envelope are biased towards radial orbits while those in the core are nearly isotropic (e.g. \citet{mwh06}). So we seek to describe the transition by choosing a DF that is biased towards radial orbits in the coarse-grained limit, but also allows  increasing isotropy. We achieve this by choosing in equations (\ref{zeroDF}) and (\ref{itP11}) respectively  
\be
K=K_o~\kappa^{\left(\frac{a-2}{a-1}\right)},
\ee
and 
\be
K'=K_1,
\ee
where $K_1$ and $K_o$ are constants. This yields 
\be
P_{oo}=K_o~{\cal E}^{1/2}/\zeta_2^2,
\label{TDF0}
\ee
and 
\be
P_{11}=-\frac{\gamma_1K_o}{2}{\cal E}^{-1/2}/\zeta_2^2+K_1~{\cal E}^{\frac{3}{2(a-1)}}.
\label{TDF1}
\ee

The motivations for this choice of DF are:

(i) It is evidently consistent with the general coarse-graining series in spherical symmetry with velocity anisotropy when stopped at second order, as we have shown above;

(ii) The equilibrium term, $P_{oo}$, is  a radially biased ${\cal E}^{1/2} $ law \citet{hw99};

(iii)  The first order correction, $P_{11}$ contains a growing isotropic term plus a radially biased ${\cal E}^{-1/2}$ law \citet{fridman}.

Thus we might expect this DF to describe a transition from primarily radial orbits in a steep density profile through the beginning of isotropic flattening. The value of the `running' self-similarity index $a$ should be greater than $1$ to describe the steep nature of this transition region.

We shall not give the calculation  for this case in detail. It suffices to affirm that we find in fact that there is no choice of parameters in this DF which allows the correction factor in the pseudo phase-space density to be small while that in  the density is substantial.  Thus we expect both of these quantities to have the power-law behaviour found by setting the left-hand sides of equations  (\ref{linear}), (\ref{1stden}) equal to spatial constants. The self-similar index $a$ should then be close to its cosmologically fixed value.   

It is  clear that if in fact we do find a region where $C$ can be small while the density is substantially flattened ($a<1$), then the power of the pseudo-density will steepen in zeroth order (see lhs of \ref{linear}) in this region for $e\ne 2/3$. We shall see however that $C$ can compensate somewhat for this zeroth order behaviour in such a region, as does the limit $e\rightarrow 2/3$.
  
We consider in detail the situation in the isotropic, flat, core in the next subsection.

\subsection{Isotropic, Flat, Core}

We turn now to the region that we know to have a flattened density and  a power-law pseudo-density \citep{dehnen05,barnes06}. We know also that this region is close to isotropic \citep{hjs,barnes05,mwh06}. Thus we should choose a DF based on eqs.(\ref{zeroDF}) and (\ref{itP11}) that has $a<1$ and is isotropic even in the first order correction term that describes the approach to the ultimate steady state. We do this by choosing $K=K_o$ and $K'=K_1$ where both $K$ and $K_1$ are constants. Hence in this section we use 
\be
P_{oo}=K_o~{\cal E}^{\frac{(3-a)}{2(a-1)}},
\label{isozeroDF}
\ee
and 
\be
P_{11}=-|\gamma_1|K_o(\frac{3-a}{2(a-1)}){\cal E}^{\left(\frac{5-3a}{2(a-1)}\right)}+K_1~{\cal E}^{\frac{3}{2(a-1)}}.
\label{isoP11}
\ee
The energy ${\cal E}$ is positive in this case, namely ${\cal E}=(\zeta_1^2+\zeta_2^2)/2+|\gamma_o|$.

We recall our earlier remarks after equation (\ref{zeroDF}) regarding the generality of the zeroth order $P_{oo}({\cal E})$. The term $P_{11}({\cal E})$ is the first order correction describing the time dependent approach to the zeroth order {\it for $R$ not too small}. 

It is now straight forward but exceedingly tedious to calculate the quantities 
$I_{oo}$, $I_{11}$, $M_{oo}$ and $M_{11}$. We will discuss only the simplest calculation here to indicate the pattern. But accuracy is of the essence and every practical check of the algebra has been made, including two independent deductions of the key result.

The calculation of $I_{oo}$ requires the evaluation of the multiple integral (for brevity we set $p\equiv (3-a)/(2(a-1))$)
\be
I_{oo}=2K_o\int_0^\infty~d\zeta_2^2~\int_{|\gamma_o|+\zeta_2^2/2}^\infty~d{\cal E}\frac{{\cal E}^p}{\sqrt{2({\cal E}-|\gamma_o|)-\zeta_2^2}},
\ee
which can be put in the form
\be
\sqrt{2}K_o\int_0^\infty~d\zeta_2^2~(|\gamma_o|+\zeta_2^2/2)^{-\frac{1}{1-a}}~\int_1^\infty~du~\frac{u^p}{\sqrt{u-1}},
\ee
and hence
\be
I_{oo}=2\sqrt{2}K_o\frac{1-a}{a}B(arg1,1/2)|\gamma_o|^{-\frac{a}{1-a}}.
\label{exIoo}
\ee
Here $B(x,y)$ is the beta function and 
\be
arg1\equiv \frac{1}{1-a}.
\label{arg1}
\ee
But by definition (\ref{zpot}) ($a\ne 2/3$) we have $I_{oo}\equiv 2(1-a)(3-2a)|\gamma_o|$ so that we obtain a relation that is much used in the other results namely
\be
|\gamma_o|^{\frac{1}{1-a}}=\frac{\sqrt{2}K_o}{a(3-2a)}B(arg1,1/2).
\label{gammao}
\ee
We note that once the index $a$ is given, the only unknown at zeroth order is $K_o$.

Proceeding steadfastly in this fashion we find to this order
\be
I_{11}=\frac{\frac{4\sqrt{2}K_1}{3}\frac{1-a}{a}|\gamma_o|^{-\frac{3a}{2(1-a)}}B(arg2,1/2)}{D_{11}},
\label{exI11}
\ee
where
\be
D_{11}\equiv 1-\frac{(3-a)(3-2a)a}{3(3a-2)(1-a)}\frac{B(arg4,1/2)}{B(arg1,1/2)}.
\label{D11}
\ee
We have set 
\be
arg2\equiv \frac{2+a}{2(1-a)},~~~~~arg4\equiv \frac{2-a}{1-a}.
\label{arg24}
\ee
Moreover provided that $a>1/2$ (otherwise the integral over $\zeta_2$ must have an arbitrary upper cut-off)
\be
M_{oo}=4\sqrt{2}K_o\frac{1-a}{2a-1}|\gamma_o|^{\frac{1-2a}{1-a}}B(arg5,3/2),
\label{exMoo}
\ee
and 
\be
arg5\equiv \frac{a}{1-a}.
\label{arg5}
\ee
Finally we may write using equation (\ref{psi1})
\bea
M_{11}&=&\frac{8\sqrt{2}(1-a)K_1}{5a-2}|\gamma_o|^{-\frac{5a-2}{2(1-a)}}B(arg3,3/2)\nonumber\\
&+ &\frac{2\sqrt{2}I_{11}K_o}{3}\frac{3-a}{a(3a-2)(1-a)}|\gamma_o|^{-\frac{a}{1-a}}B(arg1,3/2)
\label{exM11}
\eea
where
\be
arg3\equiv \frac{3a}{2(1-a)}.
\label{arg3}
\ee

All of these preliminaries allow us to write purely as a function of $a$ the ratio that appears in  $C$ as 
\bea
ratio&\equiv&\frac{I_{oo}M_{11}}{I_{11}M_{oo}}=\frac{(3-a)(3-2a)(2a-1)}{3a(3a-2)(1-a)}\frac{B(arg1,3/2)}{B(arg5,3/2)}\nonumber\\
&+ &\frac{3(2a-1)}{5a-2}\frac{B(arg1,1/2)B(arg3,3/2)}{B(arg2,1/2)B(arg5,3/2)}D_{11},
\label{ratio}
\eea
whence follows the explicit correction factor (we repeat equation \ref{1stratio} in a convenient form)
\be
C\equiv\frac{3e}{2}(ratio-1)-1.
\label{cf}
\ee

The calculation of $C$ depends only on the  self-similarity index $a$ and $e$. We are therefore in a position to ask for the function $a(e)$, which is defined by setting $C$ equal to a value $<<1$. Should this prove to be possible for a range of $e$ about unity, and for values of the running index that appear dynamically in the isotropic centre, then we might conclude that the pseudo-density with $e=1$ has no special significance. Rather it would be the isotropic DF, together with the collective relaxation that leads to $a<1$ \citep{hjs,austin05,barnes05,mwh06} that  produces a power-law pseudo-density for a variety of values of $e$ close to unity \citep{barnes06}. 

We have verified  that $C$ is never small according to this type of analysis  for an isotropic centre with $a>1$. Hence such a configuration would require the density and the pseudo-density to flatten to the same order relative to their equilibrium power laws, which is not found. This  is another indication that $a$ must decrease below unity as the centre of the halo is approached

We shall see in our subsequent discussion that nevertheless (for general $e$ and a running $a$) slight  deviations from a pure power-law behaviour for the pseudo-density are predicted by our analysis ( there may be a hint of such behaviour in \citet{graham05}, section 7). This is because the steepening relative to $-2$ produced by the declining index $a$ is not always exactly compensated by a flattening due to the first order correction.
  \citet{barnes06} however  conclude that $e=1$ is superior to neighbouring values and that it yields a strict power-law greater than $-2$ . It may be that this result is not yet conclusive (e.g.\citep{graham05}), or perhaps as we suggest in the discussion that this is a measure of incomplete relaxation in the simulations. This argument requires $e=2/3$ to form the ultimate relaxed power-law pseudo-density since then the equilibrium power-law of the pseudo-density is $-2$ independent of $a$ ,which can therefore remain fixed near a value that renders $C\approx 0$ throughout the halo. There is as yet no evidence for this state in the simulations.

.
 
\begin{figure}
\epsscale{1.0}
\plotone{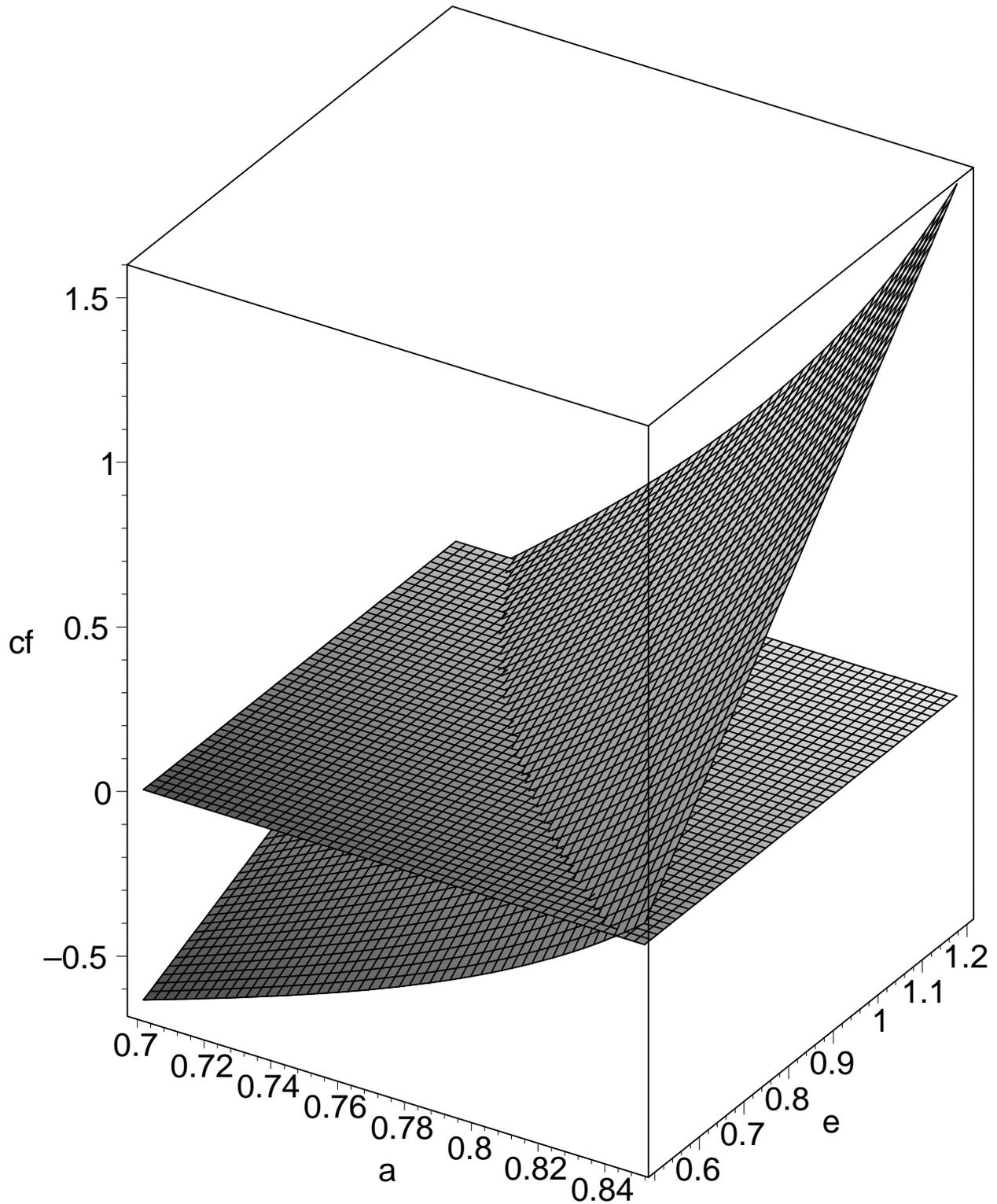}
\caption{The figure shows the correction factor C plotted as a function of $a$ and $e$. The horizontal plane corresponds to $C=0$. The intersection of the plane with the $C(a,e)$ surface defines the curve $a(e)$ for $C=0$. The case $e=1$ has a zero C for $a\approx 0.7403$, while $e=2/3$ requires $a=0.8$ almost exactly. Intermediate values are $e=0.8$, $a\approx 0.7726$; $e=1.2$,$a\approx 0.7165$.}

\label{fig:cf}
\end{figure}

In figure  (\ref{fig:cf})   $C(a,e)$ is plotted over ranges of $a$ of interest. It is evident that the scenario previewed above holds over a rather limited range in $a$. So we are only permitted to admit small variations about the critical $a(e)$ where $C=0$ in the figure. 

We see that there is a very precisely defined $a$ such that $C=0$ over a range of $e$ close to unity.  This value is broadened if, for example, we were to be satisfied with $C=0.2$. The numerical values of $2a$ required to maintain the pseudo-density power-law lie in the range $1.4$-$1.6$. 

\begin{figure}
\epsscale{1.0}
\plotone{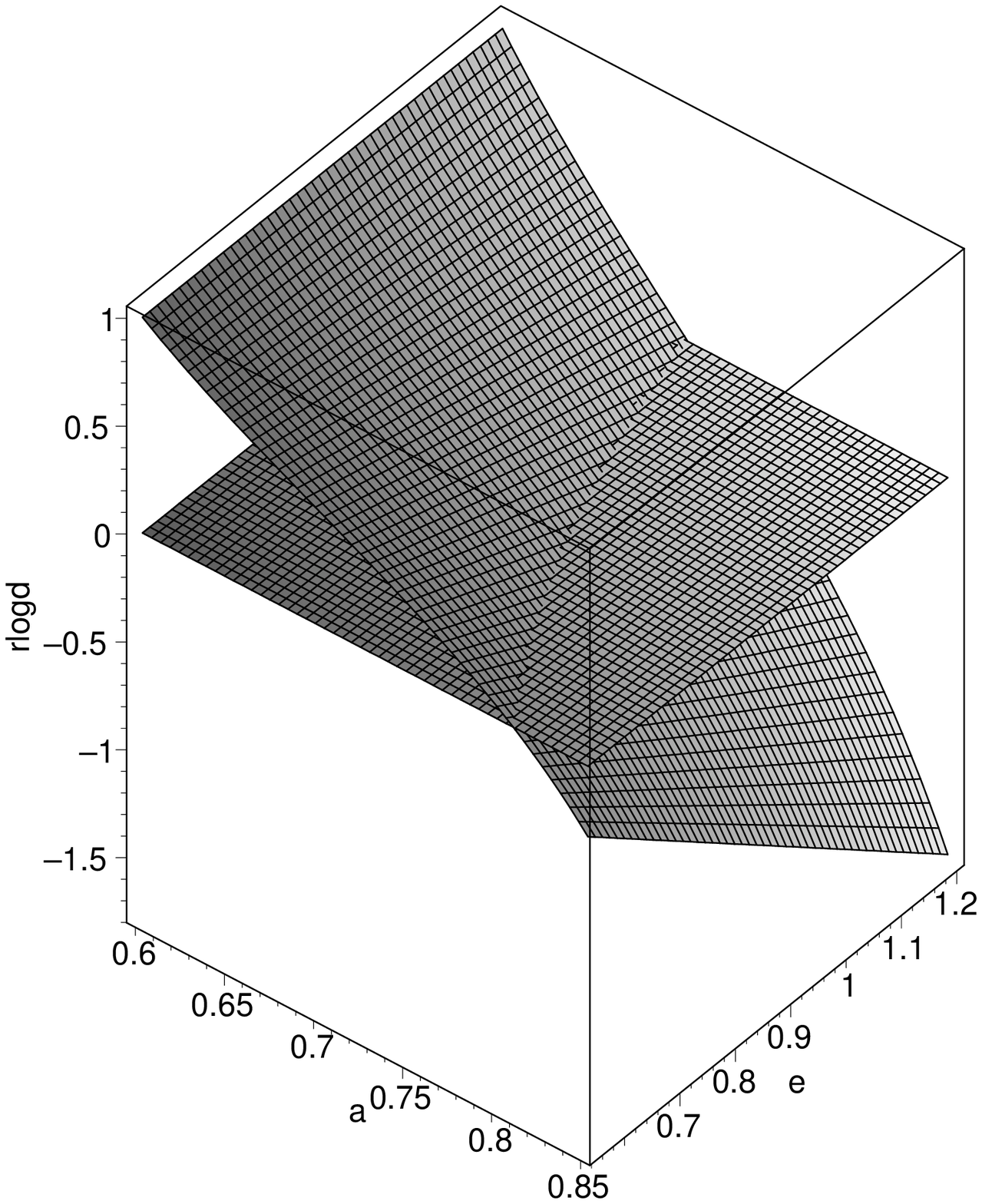}
\caption{The figure shows the ratio of the logarithmic derivatives given in equation \ref{PsDD} of the text namely $rlogds(a,e)$  plotted as a surface over $a$ and $e$ at $x=0.5$. The horizontal plane shows the zero ratio. The intersection of this plane with the surface defines the curve $a(e)$ for zero ratio, which should be compared with the condition $C=0$ in figure \ref{fig:cf}). The case $e=1$ has a zero ratio for $a\approx 0.721$, while $e=2/3$ requires $a=0.773$. Intermediate values are $e=0.8$, $a\approx 0.748$; $e=1.2$,$a\approx 0.701$.}

\label{fig:PsDD}
\end{figure}

 As a check on our linearized calculation we also present in figure (\ref{fig:PsDD}) the same calculation for the ratio of the logarithmic derivatives at a typical value of $x=0.5$. We see that numerically the results are very similar, since for $e=1$ the ratio of the logarithmic derivatives is zero at $a=0.721$ compared to $C=0$ at $a=0.740$. For $e=2/3$ the corresponding values  are $0.773$ and $0.800$. At $x=1.0$ the number for $e=1.0$ changes only to $0.728$, while at $0.1$ it becomes $0.69$. Other values of $e$ yield a similar numerical variation, but do not change the results qualitatively. We recall that the major omission from our theory is the ability to calculate $a(x)$. This might be found in a detailed fit to a simulation however.

\begin{figure}
\begin{tabular}{cc}
\rotatebox{0}{\scalebox{0.50}
        {\includegraphics
                {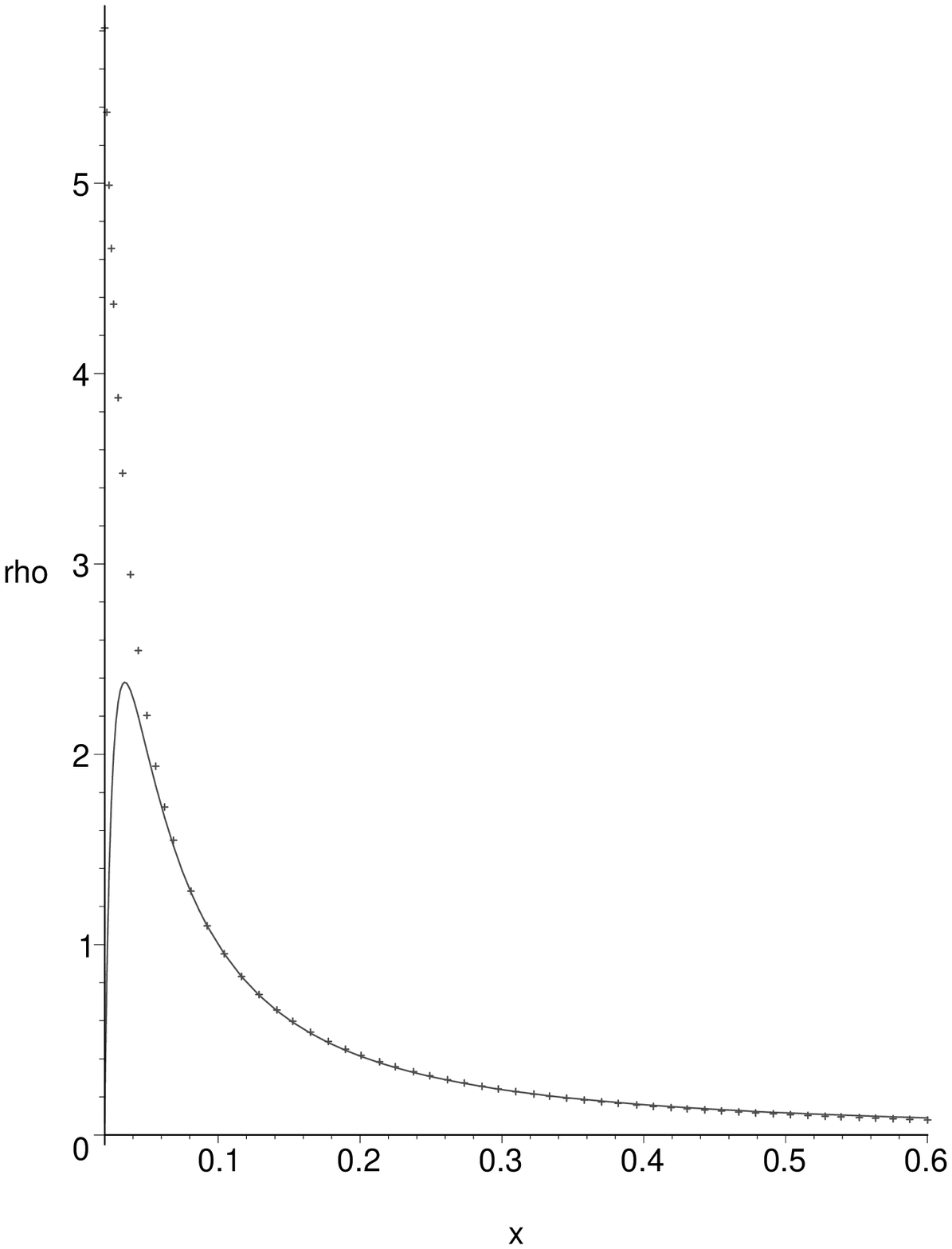}}} & 
\rotatebox{0}{\scalebox{0.50}
        {\includegraphics
                {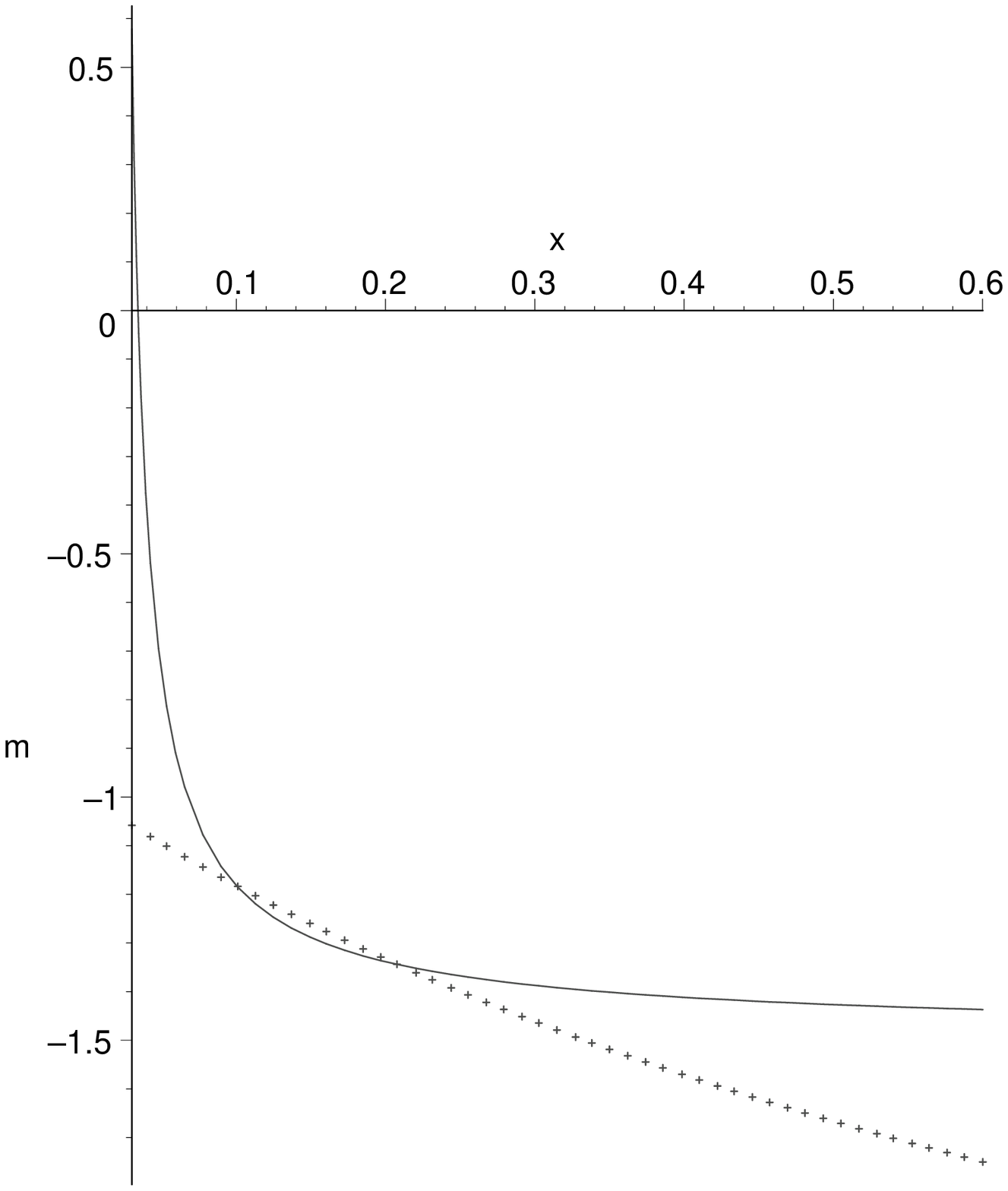}}} \\
\end{tabular}
\caption{\label{fig:curvefit}
The left box of the figure shows an example of the fit to the NFW profile (dotted) that is possible over a limited region  in radius with the first order density formula in the text (line). The profiles have been adjusted to have the same slope of $-1.1818$ and the same normalized unit value at $x\equiv r/r_o=0.1$. This requires $I_{11}/(\alpha I_{oo})\approx 0.053$. The fit has been made with $a=0.75$, which is in the middle of the range that gives zero for the pseudo-density correction. One sees that as expected the fit fails at sufficiently small $x$, but is reasonable over a limited range $>0.1$. The right box illustrates how  sensitive  the logarithmic slope is to the fitting function over this same range by plotting the logarithmic slope $m$ of the NFW function (dotted) and of the fitting function (line).} 
\end{figure}

These density profile powers are associated with the NFW central region \citep{moore98,diemand05}, but they do not correspond to the flattest density profile found in the simulations when used in zeroth order (see e.g. eq. \ref{1stden}). However we must remember the first order correction factor for the density on the rhs of eq. (\ref{1stden}). Indeed the significance of our result is that for the appropriate value of $a$, the density correction factor may be substantial while that of the pseudo-density is essentially zero. An example of the fit to the NFW density profile that is possible with the form (\ref{1stden}) is shown in figure (\ref{fig:curvefit}) when $a=0.75$, close to the critical line in figures \ref{fig:cf},\ref{fig:PsDD}). This is only illustrative, but it could be considered a reasonable fit to the simulated data  of \citet{diemand05} (their figure 11) in the range $10^{-3}r_{virial}$ to $10^{-2} r_{virial}$ (our $x$ has an arbitrary scale). The figure also shows  how  sensitive  the logarithmic slope is  to the fitted curve compared to the curve itself.  

\clearpage

We have noted that the pseudo-density powerlaw should be $-((2-3e)a+3e)\approx -2$ in terms of the index $a$ in the inner envelope (i.e. near but outside the NFW scale radius) where $a\ge 1$   Thus there is no reason to suspect that $a$ should vary much from its cosmologically fixed value (typically $9/8$) until the edge effects  near the virial radius destroy the self-similarity. We recall that near the virial radius  environmental (\citet{mwh06}) and mass exhaustion effects (\citet{hw99,lu}) can dominate the internal relaxation. As an example of typical outer behaviour we use $a=9/8$, which gives the pseudo-density power as $-(18-3e)/8$. This gives $-15/8$ for $e=1$ and $-2$ for $e=2/3$ (in the latter case the power is $-2$ strictly independently of $a$ as already remarked). These bracket simulated values.  

By way of predictable behaviour inside the scale radius we see from  equation (\ref{linear}) that the zeroth order pseudo-density steepens ($e\ne 2/3$) relative to $-2$ as $a$ decreases from $1$, which we expect to happen by the arguments in our introduction. This occurs while the density is flattening relative to $-2$ also in zeroth order, although the effect in the pseudo-density is weaker as $e\rightarrow 2/3$ .
 
The first order correction adds to the steepening for $a>a(e)$ where $C>0$ (fig. \ref{fig:cf}), but decreases to zero on the critical curve $a(e)$. 
 After passing through zero, the first-order term acts against the zeroth order steepening (figure(\ref{fig:cf}) shows $C\approx -0.61$ at $a=0.65$  for $e=1$; $rlogds\approx 0.57$ ), although it is relatively weak relative to the first order density correction (the correction factors are not equal in eqs. (\ref{1stden}) and (\ref{linear}) or (\ref{PsDD}) until $a=0.6$ for either $e=1$ or $e=0.67$).  
Very similar behaviour is predicted from the exact ratio of the logarithmic derivatives as shown in figure (\ref{fig:PsDD}) provided the density is itself less steep than $r^{-2a}$. 

Such a slope oscillation would be detectable in the simulations, but is not strongly evident \citep{dehnen05,barnes06,graham05}. The oscillation would be less pronounced at smaller $e$ and in fact vanishes at $e=2/3$ when the zeroth-order power of the pseudo-density becomes independent of $a$, but then we must remember that $C$ is not zero. It  vanishes at $a=0.8$ for this value of $e$, but as it is positive at larger $a$ and negative at smaller $a$ (figure \ref{fig:cf}), it would by itself produce an outer steepening followed by an inner flattening for this value of $e$.  We must nevertheless emphasize that this remains a truer power-law than that of the density, since coincident with this behaviour, equation (\ref{1stden}) shows that the zeroth and first order terms are combining to flatten the density profile monotonically relative to the outer power-law. They can produce
the NFW central index of $-1$ (figure \ref{fig:curvefit}) well before $a$ reaches the value $1/2$, which would by itself produce this behaviour in zeroth order. We do not expect $a$ to be so far from the critical line $a(e)$ however until complete relaxation is achieved. 

 There is one possible exception to the first order slope oscillation expected when $e=2/3$ due to the $C$ variation of figure (\ref{fig:cf}).This will be discussed further in our conclusion section, but it requires that $a$ sticks at the value $\approx 0.8$ where $C$ vanishes  for this $e$. This would produce a rigorous $r^{-2}$ power law for $\rho/\sigma^2$ throughout the halo.

To summarize based on a specific case, the  first order flattening of the zeroth order power-law $a-3$ alluded to above is active when $a$ is in the  range $]0.6,0.74]$ for $e=1$. This will tend to preserve the power-law in the pseudo-density which would otherwise steepen in zeroth order as $a$ decreases, but there could be detectable steepening as $a$ varies from about $1$ down to $0.74$ since $C$ acts also to steepen the pseudo-density in this regime being positive (\ref{fig:cf}). However if this transition in the density power happens relatively rapidly in radius, it may not be pronounced (e.g. figure 10 \citep{diemand05}, appearing only as a small `bump' on the curve.  Reducing $e$ towards $2/3$ will  remove the zero-order steepening, but makes the first order flattening about equal to that of the density at $a=0.6$ ( about one half this amount at $0.7$, but  zero at $a=0.8$- all as in figure \ref{fig:cf}). This should be measurably worse than  the mean curve for $e=1$, if $a$ really does run through this range of values, as is concluded in  current simulations \citep{barnes06}. 

The ratio of the logarithmic derivatives gives us slightly different information than the linearized treatment, but it is largely consistent with it. It incorporates the dependence on radius that proves to be relatively weak for our current purposes when the first order terms are smaller than unity. More importantly figure (\ref{fig:PsDD}) being roughly independent of position, shows that when the density is flatter than $r^{-2a}$ on the high $a$ side of the figure, the pseudo-density will be steeper than $r^{-((2-3e)a+3e)}$ and conversely on the low $a$ side.

\section{Discussion and Conclusions}

In this paper we have used two analytic ideas. One was the introduction of the running or adiabatic self-similarity, while the other was  the series coarse-graining of the self-similarity which allows a description of the approach to the final steady state due to internal relaxation. It would in principle predict the relaxation in non-spherical symmetry, particularly if a renormalization improvement were found. 

We have used the second technique to deduce a DF to zeroth and first order that can apply in the flat density, isotropic, limit. Such is expected to be the case inside the NFW scaling radius of a halo. This DF appears quite naturally in many places \citep{Tretal, hw95} in zeroth order. We used this DF to calculate the first order corrections to the density and to the phase-space pseudo-density (generalized slightly by the parameter $e$). We have found a range of the running index $a(e)$ over which the first order correction to the density may be substantial while that of the pseudo-density is negligible. 

More generally, the transitory first order term together with the running index exert compensating effects that can maintain the slope of the pseudo-density sensibly constant over a wider range of $a$. This is not expected to change until, and if, a rigorously flat core is detected in the simulations. The special case where $e=2/3$, so that we consider in fact the function $\rho/\sigma^2$, should show a very good power-law with slope $-2$ between the virial radius and the scale radius, after  which the slope will show an oscillation (ultimately flattening ) relative to this power-law. 

The region of predominatly radial orbits between the NFW scale radius and the virial radius (ignoring edge effects) should be described accurately by the original radial in-fall models \citep{fg84,bert85}, which predict in our terms a slope of $-((2-3e)a+3e)$ for the pseudo phase-space density. This is  $-2$ for $a=1$, and $-15/8$ for $a=9/8$ and $e=1$. It is also $-2$ for $e=2/3$.

It is important to note that not just any DF allows the pseudo-density to be a power-law while the density flattens. In particular neither a steep density, isotropic DF, nor a radially biased, steep density DF does so. Neither the density nor the pseudo-density flattens markedly in the envelope where the latter DF is relevant, if edge effects are ignored (e.g. \citep{mwh06}).
 
Thus we conclude that the remarkable power-law behaviour in the pseudo-density \citep{taylor01} and its related forms is a consequence of the isotropic DF produced in the core by  processes of relaxation, such as the radial orbit instability \citep{pol81,me85}  . Because the behaviour is found to be similar for a reasonable range of $e$ about $e=1$ (especially on the low side) we do not think that the form $\rho/\sigma^3$ is especially significant (see also \citet{hansen04}, \citet{dehnen05}, \citet{barnes06}). It should be admitted however that currently most simulations \citep{dehnen05, barnes06} do reveal $e=1$ as the superior power law although other values of $e$ produce power laws superior to that of the density (e.g. \citep{barnes06}) as we would expect. Should there prove ultimately to be a  preference for $e=1$, it would remain a puzzle according to the present analysis. It might be attributed to higher order behaviour since after all this is a first order calculation. We give an alternate suggestion in our speculation below.

The approximate constancy of the NFW concentration is guaranteed in our treatment only by the fact that the evolution is taken to be adiabatically self-similar throughout. Together with the previous paragraph, we have thus answered in the affirmative, at least tentatively, the questions posed in the introduction. 
 
It seems that we have not penetrated very far into the details of the underlying relaxation however, beyond suggesting that the equations do know about it! But we note that our explanation of the power-law behaviour in the pseudo phase-space density invokes a running or adiabatic self-similarity. This is thought to arise as collective effects change the dominant conserved quantities from those set by initial conditions to those compatible with an isotropic distribution function and an isolated core \citep{henrik06}. The radial orbit instability is one collective effect that achieves this \citep{hjs,barnes05, mwh06}, but in spherical symmetry we can only imitate such effects by using appropriate DF's at the beginning and end of the relaxation process.

Finally on a speculative note, we observe that  
$\rho/\sigma^2$ is predicted here to have a slope rigorously $-2$ until significantly inside the NFW scale radius. Such a relation implies that the local Jeans' length (ignoring the difference between the total dispersion and the radial dispersion for the moment) scales linearly with radius. Moreover, as discussed in \citep{henrik06}, we might expect the collective relaxation to proceed by Landau-damping on waves created below or at this scale. Hence this relation states that the relaxation is effective up to a fixed fraction, probably close to unity, of the local scale. This would be perceived as violent `virialization' on the large scale and asymptotic `thermalization' on the small scale.

 The flattening deviation from the rigorous inverse square behaviour of this quantity would  indicate the absence of such relaxation (the Jeans' length becomes greater than the local scale). 
The apparent superiority of $e=1$ in the current simulations may be due to the expected radial variation of $\sigma\propto r^{(1-a)}$. Thus while $\rho/\sigma^2$ flattens in the absence of complete relaxation, dividing by an additional factor $\sigma$ at small $r$ for $a<1$ acts against this flattening since $\sigma$ is small there. We would predict on these grounds that in simulations with higher resolution $\rho/\sigma^2$ should become a better power law than $\rho/\sigma^3$. 

On this view the essential physics is in the behaviour of $\rho/\sigma^2$ and the physical running value of $a$ would be set by setting $C=0$ for $e=2/3$. This suggests that the physical core value of $a$ would be $\approx 0.8$ in a perfect simulation.  Subsequent flattening of the density would be due to first and higher order terms. Fig.(\ref{fig:curvefit}) is not changed essentially by using $a=0.8$, so that such flattening is at least possible in a transitory fashion. Ultimately the renormalization group approach used by \citet{McD06} may allow a longer lived description.


\acknowledgements{
This work was supported in part by the canadian Natural Science and Engineering Research Council. I wish to thank colleagues and particularly St\'ephane Courteau for a careful reading of the manuscript. An anonymous referee has greatly improved the  presentation of this work by constructive criticism.}



\newpage

\begin{thebibliography}{}

\bibitem[Austin \emp{et al} (2005)]{austin05} Austin, C.G., Williams, L.R., Barnes, E.I., Babul, A., \& Dalcanton, J.J. 2005,\apj,634,756




\bibitem[Barnes \emp{et al.}(2005a)]{barnes05} Barnes, E.I., Williams,
L.L.R., Babul, A., \& Dalcanton, J.J 2005, ApJ, 634, 775

\bibitem[Barnes \emp{et al.}(2006)]{barnes06} Barnes, E.I., Williams,
L.L.R., Babul, A., \& Dalcanton, J.J. 2006, ApJ,643, 797

\bibitem[Bertschinger(1985)]{bert85} Bertschinger, E.  1985, ApJS, 58,
39




\bibitem[Carter \& Henriksen(1991)]{CH91} Carter, B., \& Henriksen, R. N. 1991,
J.Math.Phys., 32, 2580








\bibitem[Dehnen \& McLaughlin(2005)]{dehnen05} Dehnen, W., \&
McLaughlin, D.E.  2005, MNRAS, 363, 1057

\bibitem[Diemand  \emp{et al.} (2005)]{diemand05} Diemand, J., Zemp,M., Moore, B., Stadel, J., \& Carollo, M. 2005, MNRAS, 364, 665
 

\bibitem[Evans \& Collett(1997)]{EC97} Evans, N. W., \& Collett, J.L. 1997, \apj, 480, L103

\bibitem[Fillmore \& Goldreich(1984)]{fg84} Fillmore, J.A., \&
Goldreich, P.  1984, ApJ, 281, 1



\bibitem[Fridman \& Polyachenko(1984)]{fridman} Fridman, A. M. \&
Polyachenko, V. L. 1984, {\it Physics of Gravitating Systems}
(New York: Springer)



\bibitem[Graham et al.(2005)]{graham05} Graham, A.W., Merritt, D., Moore, B., Diemand,J. \& Terzi\'c, B., 2005, astro-ph/0509417



\bibitem[Hansen(2004)]{hansen04} Hansen, S. 2004, MNRAS, 352, L41
2005


\bibitem[Henriksen(1997)]{henrik97} Henriksen, R.N. 1997,in \lq Scale Invariance and Beyond'', Dubrulle, B., Graner, F.\& Sornette,  D.,(eds), Les Houches Workshop, Springer, Berlin,63

\bibitem[Henriksen \& Widrow(1995)]{hw95} Henriksen, R.N., \& Widrow,
L.M.  1995, \mnras, 276, 679

\bibitem[Henriksen \& Widrow(1997)]{hw97} Henriksen, R.N., \& Widrow,
L.M.  1997, \prl, 78, 3426

\bibitem[Henriksen \& Widrow(1999)]{hw99} Henriksen, R.N., \& Widrow,
L.M.  1999, MNRAS, 302, 321

\bibitem[Henriksen \& Le Delliou(2002)]{hld02} Henriksen, R.N., \& Le Delliou,
M. 2002, \mnras, 331,423

\bibitem[Henriksen(2004)]{henrik04} Henriksen, R.N. 2004, MNRAS, 355, 1217

\bibitem[Henriksen(2006)]{henrik06} Henriksen, R.N. 2006, MNRAS, 366, 697


\bibitem[Huss, Jain, \& Steinmetz(1999)]{hjs} Huss, A., Jain, B., \&
Steinmetz, M.  1999, ApJ, 517, 64


\bibitem[Kulessa \& Lynden-Bell(1992)]{kl92} Kulessa, A.S.\& Lynden-Bell, D. 1992,\mnras,  255, 105


\bibitem[Lu \emp{et al.}(2006)]{lu} Lu, Y., Mo, H. J., Katz, N.,
\& Weinberg, M. D. 2006, \mnras, 368, 1931


\bibitem[MacMillan, Widrow \& Henriksen(2006)]{mwh06} MacMillan, J.D., Widrow, L.M. \&
Henriksen, R.N., 2006, \apj, in press


\bibitem[Merritt\& Aguilar(1985)]{me85} Merritt, D. \& Aguilar, L. A., 1985, \mnras,217,787

\bibitem[McDonald (2006)]{McD06} McDonald, P. 2006, astro-ph 0606028

\bibitem[Moore \emp{et al.}(1998)]{moore98} Moore, B., Governato, F.,
Quinn, T., Stadel, J., \& Lake, G.  1998, ApJ, 499, L5

\bibitem[Navarro, Frenk, \& White(1997)]{nfw} (NFW) Navarro, J.F.,
Frenk, C.S., \& White, S.D.M.  1997, ApJ, 490, 493




\bibitem[Peacock(1999)]{peacock99} Peacock, J. 1999, {\it Cosmological Physics},
Cambridge University Press, Cambridge, U.K.


\bibitem[Peebles(1993)]{peebles93} Peebles, P.J. E. 1993, {\it Principles of 
Physical Cosmology}, Princeton University Press, Princeton, NJ

\bibitem[Polyachenko(1981)]{pol81} Polyachenko, V.L., 1981, Soviet Astr.Let., 7, 79








\bibitem[Taylor \& Navarro(2001)]{taylor01} Taylor, J.E., \& Navarro,
J.F.  2001, ApJ, 563, 483

\bibitem[Tremaine et al(1994)]{Tretal} Tremaine, S., Richstone, D.O., Yong-Ik Byun, Dressler, A., Faber,S.M., Grillmair, C., Kormendy, J.\& Lauer, T.R. 1994,\aj, 107,634




\end{thebibliography}
\end{document}